\documentstyle[12pt]{article}
\begin{document}
\newcommand{\nd}{\noindent}
\newcommand{\beq}{\begin{equation}}
\newcommand{\eeq}{\end{equation}}
\newcommand{\barr}{\begin{eqnarray}}
\newcommand{\earr}{\end{eqnarray}}
\newcommand{\NP}[1]{{\it Nucl.\ Phys.}\ {\bf #1}}
\newcommand{\ZP}[1]{{\it Z.\ Phys.}\ {\bf #1}}
\newcommand{\PL}[1]{{\it Phys.\ Lett.}\ {\bf #1}}
\newcommand{\PR}[1]{{\it Phys.\ Rev.}\ {\bf #1}}
\newcommand{\PRL}[1]{{\it Phys.\ Rev.\ Lett.}\ {\bf #1}}
\newcommand{\MPL}[1]{{\it Mod.\ Phys.\ Lett.}\ {\bf #1}}
\newcommand{\SNP}[1]{{\it Sov.\ J.\ Nucl.\ Phys.}\ {\bf #1}}
\def\lsim{\mathrel{\rlap{\lower4pt\hbox{\hskip1pt$\sim$}}
    \raise1pt\hbox{$<$}}}         
\def\gsim{\mathrel{\rlap{\lower4pt\hbox{\hskip1pt$\sim$}}
    \raise1pt\hbox{$>$}}}         
\def\kp{\mbox{\boldmath $k_\perp$}}
\def\bp{\mbox{\boldmath $p$}}
\begin{flushright}
DFTT 65/94 \\
hep-ph/9412313 \\
October 1994
\end{flushright}

\vspace{0.2in}

\renewcommand{\thefootnote}{\fnsymbol{footnote}}

\begin{center}
{\LARGE Spin effects in strong interactions at high energy%
\footnote{Rapporteur talk at the XI International
Symposium on High Energy Spin Physics, September 15-22,
Bloomington, Indiana, to be published in the Proceedings}
}
\end{center}

\vskip8pt

\begin{center}
{\large M. Anselmino}
\end{center}

\vskip8pt

\begin{center}
{\small\it Dipartimento di Fisica Teorica, Universit\`a
di Torino and \\
Istituto Nazionale di Fisica Nucleare, Sezione di Torino, \\
Via P. Giuria 1, I-10125 Torino, Italy }
\end{center}

\vskip24pt

\begin{center}
\nd \parbox{5.0in}{\small {\bf Abstract}
Spin effects in strong interaction high energy processes are subtle phenomena
which involve both short and long distance physics and test perturbative and
non perturbative aspects of QCD. Moreover, depending on quantities like
interferences between different amplitudes and relative phases, spin
observables
always test a theory at a fundamental quantum mechanical level; it is then no
surprise that spin data are often difficult to accomodate within the existing
models. A report is made on the main issues and contributions discussed in
the parallel Session on the ``Strong interactions at high energy" in this
Conference.}
\end{center}

\vskip12pt
In the parallel Session on ``Strong interactions at high energy" a total of 22
talks were presented on several and various subjects, both theoretical and
experimental. As usual, spin effects prove to be unexpected, rich and difficult
to understand, testing the deep and basic properties of any theory. In high
energy interactions the underlying fundamental theory we are testing is QCD,
both in its perturbative and non perturbative aspects; the latter, at least in
the energy region so far explored, still play a crucial role, as has emerged
from most of the presentations.

Rather then considering the single contributions I have selected
the main topics and experimental or theoretical problems discussed in the
Session and will try to comment and report on these issues. For more details
on the single contributions one is referred to the original talks published
in these Proceedings. I apologize with some authors whose contributions will
receive less attention than others; this is only due to my organization of the
material and by no means implies a negative judgement on their work.

The arguments most widely discussed can be divided into: \\\nd
1) $pp$ elastic scattering and total cross-sections; \\\nd
2) single spin inclusive asymmetries; \\\nd
3) hyperon polarization; \\\nd
4) other topics; \\\nd
5) nucleon spin crisis.

The last subject was actually extensively treated in the parallel Session
on ``Nucleon spin structure functions" to which one is referred; I shall
only make here a short comment to answer many questions raised at various
stages during the Conference.

\vskip 6pt
\nd
{\bf \bp \bp \,\, elastic scattering and total cross-sections}
\vskip 4pt
Surprising spin effects in large angle proton-proton elastic scattering at
high energy have been known for quite a long time \cite{ssa,dsa}: they concern
both single and double spin asymmetries, $A, A_{NN}, A_{LL}$ and $A_{SS}$.
Consider the C.M. scattering of two protons moving along the $z$-axis and
let $(xz)$ be the scattering plane; then the single spin asymmetry $A$, or
analyzing power, is defined as
\beq
A\equiv {d\sigma^\uparrow - d\sigma^\downarrow \over
d\sigma^\uparrow + d\sigma^\downarrow}
\eeq
where $d\sigma^{\uparrow(\downarrow)}$ is the cross-section for the elastic
scattering of an unpolarized proton off a proton polarized perpendicularly
to the scattering plane, that is along ($\uparrow$) or opposite ($\downarrow$)
the $\hat y$-axis.

The double spin asymmetry $A_{NN}$ is defined by
\beq
A_{NN} \equiv {d\sigma^{\uparrow\uparrow} - d\sigma^{\uparrow \downarrow}
\over d\sigma^{\uparrow\uparrow} + d\sigma^{\uparrow \downarrow}}
\eeq
where, now, both initial protons are polarized in the direction normal
($N$) to the scattering plane, as explained for $A$.

Analogously to $A_{NN}$, one can define the double spin asymmetries
$A_{SS}$ and $A_{LL}$ which only differ from $A_{NN}$ by the direction
along which the initial protons are polarized: the $S$ direction is the
$\hat x$ direction, always choosing ($xz$) as the scattering plane, and
the $L$ direction is along $\hat z$.

$A, A_{NN}, A_{SS}$ and $A_{LL}$ can be written in terms of the five
independent helicity amplitudes
\barr
\phi_1 &\equiv& \langle ++|\phi|++ \rangle \nonumber\\
\phi_2 &\equiv& \langle ++|\phi|-- \rangle \nonumber\\
\phi_3 &\equiv& \langle +-|\phi|+- \rangle \\
\phi_4 &\equiv& \langle +-|\phi|-+ \rangle \nonumber\\
\phi_5 &\equiv& \langle ++|\phi|+- \rangle \nonumber
\earr
as
\barr
A &=& \Sigma^{-1}{\mbox{\rm Im}}\,
      [\Phi_5^*(\Phi_4-\Phi_1-\Phi_2-\Phi_3)] \nonumber\\
A_{NN} &=& \Sigma^{-1}{\mbox{\rm Re}}\,
      [\Phi_1\Phi_2^*-\Phi_3\Phi_4^*+2|\Phi_5|^2]\nonumber\\
A_{SS} &=& \Sigma^{-1}{\mbox{\rm Re}}\,
      [\Phi_1\Phi_2^*+\Phi_3\Phi_4^*] \\
2A_{LL} &=& \Sigma^{-1}
      [|\Phi_3|^2+|\Phi_4|^2-|\Phi_1|^2-|\Phi_2|^2] \nonumber\\
2\Sigma &=& [|\Phi_1|^2+|\Phi_2|^2+|\Phi_3|^2+|\Phi_4|^2+4|\Phi_5|^2]\nonumber
\earr

According to the standard QCD description of large angle exclusive reactions
$AB \to CD$ \cite{bl}, assuming the dominance of collinear valence quark
configurations in the proton wave function, one has the helicity conservation
rule
\beq
\lambda_A + \lambda_B = \lambda_C + \lambda_D
\eeq
which immediately implies, for $pp$ elastic scattering, $\phi_2=\phi_5=0$ and
consequently, via Eqs. (4),
\beq
A=0 \qquad\qquad A_{NN} = -A_{SS}\,.
\eeq
The experimental data \cite{ssa,dsa} do not agree with such conclusions.

All this has been known for quite a long time and several models and attempts
to explain the data can be found in the literature. In summary they amount to
introduce non perturbative contributions which should still be important at the
energies of the performed experiments; quark masses, higher order corrections
or intrinsic transverse momenta, which violate the helicity conservation rule,
ought to be properly taken into account. Also possible lower energy mechanisms
might still interfere with the hard scattering picture and
give a sizeable contribution. However, one expects that at higher energies
the perturbative QCD predictions should eventually turn out to be
correct%
\footnote{\nd This need not be true for high energy {\it small angle}
processes, where non perturbative anomalous contributions might persist and
remain energy independent in polarization measurements \cite{for}}
and further experimental information would be of great importance.

A rich and interesting program for measuring several spin observables in $pp$
elastic scattering and inclusive production is in progress at UNK (Protvino)
and RHIC (Brokhaven); reports on the status of the planned experiments can be
found in the talks by V. Solovianov, A.M.T. Lin and P.A. Draper.

On the theoretical side it was pointed out in the talk by G. Ramsey how a
careful analysis of the data on $A_{NN}$ at the C.M. scattering angle
$\theta=\pi/2$, implemented with some reasonable assumptions, might supply
information on the scattering amplitudes. $\theta=\pi/2$ is an interesting
region in that the data on $A_{NN}$ present a rich structure as a function of
the energy and the number of independent amplitudes is reduced to three
($\phi_5(\pi/2)=0\,, \phi_3(\pi/2)=-\phi_5(\pi/2)$); moreover, one can assume
the double-flip helicity amplitude $\phi_2 \simeq 0$.

S. Troshin has presented
a model which attempts to explain spin effects via the interference between
two phases of QCD, a non perturbative one, described by an effective Lagrangian
which allows helicity flips, and the usual perturbative one; such interference
is significant in the intermediate energy region of the existing data.

Spin effects are also expected in the measurements of
$\Delta\sigma_L(pp) =
\sigma_{tot}^{\to\to} - \sigma_{tot}^{\to\leftarrow}$
and $\Delta\sigma_N(pp) =
\sigma_{tot}^{\uparrow \uparrow} - \sigma_{tot}^{\uparrow\downarrow}$,
the differences between proton-proton total cross-sections in pure lonfitudinal
and normal spin states respectively. $\Delta\sigma_L(pp)$ should be sensitive
to the gluon polarization inside the proton ($\Delta g$) and any information on
$\Delta g$ is of crucial importance to settle the nucleon spin crisis issue.
Some preliminary results on this quantity and, for the first time, on
$\Delta\sigma_L(\bar pp)$ were reported by D. Grosnick; however, these results
are small, consistent with zero, with large errors, and do not allow yet to
discriminate between different theoretical models.

\vskip 6pt
\nd
{\bf Single spin inclusive asymmetries}
\vskip 4pt
Large single spin asymmetries have also been observed \cite{pias} in the
inclusive production of pions in the collision of a high energy polarized
proton beam on an unpolarized target, $p^\uparrow + p \to \pi + X$, where the
proton spin is up or down with respect to the scattering plane.
These asymmetries
\beq
A_N(x_F,p_T) = {d\sigma^\uparrow - d\sigma^\downarrow
\over d\sigma^\uparrow + d\sigma^\downarrow}
\eeq
are found to be large for the transverse momentum of the pion in the range
($0.7 \lsim p_T \lsim 2$) GeV/c and large values of $x$-Feynman ($x_F$);
actually, $|A_N|$ increases with $x_F$. $A_N$ also shows interesting isospin
dependences ($A_N^{\pi^+} \approx -A_N^{\pi^-}$).

Such results are unexpected because a naive generalization of the
QCD-factorization theorem suggests that the single spin asymmetry can be
written qualitatively as:
\beq
A_N  \sim \sum_{ab \to cd} \Delta_T G_{a/p} \otimes G_{b/p}
\otimes \hat a_N \hat \sigma_{ab\to cd} \otimes D_{\pi/c}
\label{qual}
\eeq
where $G_{a/p}$ is the parton distribution function, that is
the number density of partons $a$ inside the proton, and
$\Delta_T G_{a/p} = G_{a^{\uparrow}/p^{\uparrow}} -
G_{a^{\downarrow}/p^{\uparrow}}$ is the difference between the number
density of partons $a$ with spin $\uparrow$ in a proton with spin $\uparrow$
and the number density of partons $a$ with spin $\downarrow$ in a proton
with spin $\uparrow$; $D_{\pi/c}$ is the number density of pions resulting
from the fragmentation of parton $c$; $\hat a_N$ is the single spin
asymmetry relative to the $a^{\uparrow}b \to cd$ elementary process and
$\hat \sigma$ is the cross-section for such process.

The usual argument is then that the asymmetry (\ref{qual}) is bound to be
very small because $\hat a_N \sim \alpha_s m_q /\sqrt s$ where $m_q$ is
the quark mass. This originated the widespread opinion that single spin
asymmetries are essentially zero in perturbative QCD.

However, as it has been discussed by several authors \cite{siv}-\cite{artru},
this conclusion need not be true because subtle spin effects might modify
Eq.~(\ref{qual}). Such modifications should take into account the parton
transverse motion, higher twist contributions and possibly non perturbative
effects hidden in the spin dependent distribution and fragmentation functions.

An explicit model \cite{meng} which takes into account the orbital motion of
quarks inside polarized protons or antiprotons and describes meson formation
via $q\bar q$ annihilation, has been presented by T. Meng, who advocates
the use of non perturbative mechanisms to describe spin effects. In this model
the different sign of $A_N^{\pi^+}$ and $A_N^{\pi^-}$ simply originates from
the $SU(6)$ valence quark spin content of the protons; the results are in
qualitative agreement with data and predictions are given for the single spin
asymmetry in Drell-Yan processes $\bar p^\uparrow + p \to l^+ l^- + X$.

An approach closer to perturbative QCD has been discussed by F. Murgia;
application of the QCD-factorization theorem in the helicity basis (for which
it has been proved) modifies Eq. (\ref{qual}) into \cite{siv}
\beq
A_N  \sim \sum_{ab \to cd} I_{+-}^{a/p} \otimes G_{b/p}
\otimes \hat \sigma_{ab\to cd} \otimes D_{\pi/c}
\label{nqual}
\eeq
with
\beq
I_{+-}^{a/p}(x_F, \kp) = \sum_h [G_{h/p}(x_F, \kp) - G_{h/p}(x_F, -\kp)]
\label{defi}
\eeq
where $h$ is a helicity index. $I_{+-}$ is a new phenomenological function
of relevance for spin observables, analogous to $G_{a/p}(x_F)$ in the
unpolarized case; notice that it vanishes when $k_\perp = 0$. A simple model
for $I_{+-}$ yields results in very good agreement with the data \cite{noi}.

This approach has been criticized by Collins \cite{col} on the ground that
QCD time-reversal invariance should forbid a non zero value of Eq.
(\ref{defi});
he suggests a similar effects in the parton fragmentation process rather
than in the parton distribution functions. His criticism is discussed
in Ref. \cite{noi}.

Further theoretical work on spin asymmetries can be found in the talks
by G.J. Musulmanbekov, M.P. Chavleishvili and, together with a general
analysis of twist-3 single spin observables, by O. Teryaev.

W. Novak has presented a proposal to perform spin measurements at HERA by
inserting a polarized target into the unpolarized proton beam. N.~Saito has
shown the first results on the single spin asymmetry for direct photon
production in $p^\uparrow + p \to \gamma + X$ processes at Fermilab: the
results are small, consistent with zero within the large errors, in agreement
with the theoretical prediction of Ref. \cite{qiu}.

\vskip 6pt
\nd
\goodbreak
{\bf Hyperon polarization}
\nobreak
\vskip 4pt
Hyperon polarization is yet another example of puzzling single spin asymmetry
in inclusive processes (mainly $p + N \to H^\uparrow + X$, with unpolarized
beam and target); however, it is so well known and so much studied
(experimentally) that it deserves attention in its own. Only experimental
contributions on the subject have been presented during the Conference; when
adding this new information to the previously available one it will become
clear why no theoretical interpretation of the data has been attempted.

With some idealization and optimism the bulk of data on hyperon polarization
at our disposal up to some time ago could be summarized as:\\\nd
- all hyperons (with exception of $\Omega$) produced in the proton
  fragmentation \\\nd
  \phantom{-} region are polarized perpendicularly to the production plane;
  antihyperons \\\nd
  \phantom{-} are not polarized;\\\nd
- the $\Lambda^0$ polarization is negative and opposite to the $\Sigma$; \\\nd
- the $\Lambda^0$ polarization is almost energy independent;\\\nd
- the magnitude of the polarization increases with $p_T$ up to $p_T \simeq
  1$ GeV/c;\\\nd
- the magnitude of the polarization is independent of $p_T$ above 1 GeV/c and
  \\\nd \phantom{-} increases linearly with $x_F$;\\\nd
- the polarization depends weakly on the target type ($A^\alpha$ dependence).

Even at such ``simple" stage there was no clear theoretical explanation of the
data: some semi-classical models could explain some of the features \cite{lam}
and some other attempts were made which either related the $\Lambda$
polarization observed in $pp \to \Lambda X$ to that observed in $\pi p \to
\Lambda K$ \cite{sof} or used triple Regge models \cite{pre}. However, there is
no fundamental explanation of the old set of data and it is very hard to
understand how polarized strange quarks, which should be present inside
polarized hyperons, can be created in the scattering of unpolarized hadrons.

The situation is now even more hopeless. These are the new data, from Fermilab
E761 Collaboration, as summarized by S. Timm:\\\nd
- antiparticle polarization: anti-$\Sigma^+$ and  anti-$\Xi^-$ are polarized,
  anti-$\Lambda$ is not;\\\nd
- energy dependence: $\Sigma^+$ polarization decreases with energy,
  $\Xi^-$ polarization \\\nd
  \phantom{-} increases (in magnitude) and $\Lambda$ polarization
  remains constant;\\\nd
- $p_T$ dependence: $\Sigma^+$ polarization decreases above $p_T \simeq 1$
  GeV/c;\\\nd
- $x_F$ dependence: $\Sigma^+$ polarization increases with $x_F$, with a $p_T$
  dependent \\\nd
  \phantom{-} slope, $\Lambda$ polarization increases (in magnitude) with a
  fixed slope and $\Xi^-$ \\\nd
  \phantom{-} polarization is independent of $x_F$ for $x_F \gsim 0.4$\,.

No kind of regular pattern emerges from the data. It is then clear how a
successful description of these experimental results cannot originate from
general features of the underlying dynamics, but has to rely on subtle
non perturbative effects in the hadronization process for each single hyperon.

In the attempt of better understanding the basic mechanisms at work in the
quark recombination process the Fermilab E800 Collaboration has measured the
$\Xi^-$ and $\Omega^-$ polarization from polarized and unpolarized neutral
hyperon beams. The results, presented by K.A. Johns, are, once more, partially
expected and partially unexpected.

\vskip 6pt
\nd
\goodbreak
{\bf Other topics}
\vskip 4pt
\nobreak
I'll just mention some other talks which dealt with different subjects.
The polarization of the $J/\psi$ in pion-nucleus collisions was discussed
within perturbative QCD by W. Tang and some spin effects in the production
and decay of $D^{*+}$ and $\Lambda_c^+$ were studied by R. Rylko. Also a
measurement of the $\Omega^-$ magnetic moment by the E800 Collaboration was
shown by P.M. Border.

A. Brandenburg noticed how the angular distribution of leptons produced in
$\pi^-p \to \mu^+\mu^- + X$ processes is in disagreement with the standard QCD
parton model. A coherent treatment of $q\pi^-$ scattering succeeds in
accounting
for the data and favours the QCD sum rule pion wave function. A.P. Contagouris
has presented a complete computation of higher order corrections for Drell-Yan
processes with transversely polarized protons, $p^\uparrow p^\uparrow \to
l^+l^- + X$ and for $e^+e^- \to \Lambda^\uparrow \bar\Lambda^\uparrow$.

A simulation for the production of $W^{\pm}$ and $Z_0$ in the collision of
high energy polarized protons at RHIC, with the aim of measuring the sea
polarization, has been discussed by V. Rykov.

\vskip 6pt
\nd
{\bf Nucleon spin crisis}
\vskip 4pt
The nucleon spin structure functions have been discussed in the dedicated
parallel Session; however, as many questions were
raised during the talks regarding the present situation with the ``spin
crisis'', I'll briefly comment on that. A comprehensive review on the subject
can be found in Ref. \cite{rev}.

Recall that the measurement of $\Gamma_1^p \equiv \int_0^1 dx~g_1^p(x)$,
combined with information from hyperon $\beta$-decays, allows to obtain a value
of $a_0$, defined by
\beq
\langle P,S|\bar\psi \gamma_\mu \gamma_5 \psi|P,S \rangle = 2 m_N a_0 S_\mu\,,
\eeq
{\it i.e.} the expectation value between proton states (of four momentum $P$
and covariant spin vector $S$) of the flavour singlet axial vector current
$J^0_{5\mu} \equiv \bar\psi \gamma_\mu \gamma_5 \psi$.

In the naive parton model (with free quarks) one has
\beq
a_0 = \sum_{q,\bar q} \Delta q \equiv \Delta \Sigma = 2 S_z^q
\eeq
where $S_z^q$ is the total $z$-component of the spin carried by the quarks;
then one expects $a_0 \simeq 1$. Instead, the EMC data \cite{emc} yield,
at $\langle Q^2 \rangle \simeq 10$ (GeV/c)$^2$,
\beq
A_0 = 0.06 \pm 0.12 \pm 0.17\,.
\eeq
Such result, consistent with zero and certainly $\not= 1$, originated the
``spin crisis'' in the parton model.

However, due to the axial anomaly, we know that $J^0_{5\mu}$ is not a conserved
current,
\beq
\partial_\mu J^{0\mu}_5 = {\alpha_s \over 4\pi} \, G^a_{\mu\nu}
\widetilde G_a^{\mu\nu} \,,
\eeq
so that the expectation value of $J^0_{5\mu}$, $\langle P,S|J^0_{5\mu}|P,S
\rangle$, is scale dependent. Then $a_0$ is not forbidden to be small at
$\langle Q^2 \rangle \simeq 10$ (GeV/c)$^2$ and it might be close to 1
at smaller $Q^2$ values.

Moreover, this reflects into the fact that there is a gluonic
contribution to $a_0$, so that what we actually measure at large $Q^2$
is
\beq
a_0(Q^2) = \Delta\Sigma - 3\,{\alpha_s(Q^2) \over 2\pi} \Delta g(Q^2)
\label{ga}
\eeq
where $\Delta g$ is the spin carried by the gluons. Notice that the gluon
contribution $\alpha_s \Delta g$ does not vanish when $Q^2 \to \infty$ as the
decrese of $\alpha_s$ is compensated by an increase of $\Delta g$.
Eq. (\ref{ga}) tells us that $a_0 \simeq 0$ does not necessarily imply
$\Delta \Sigma \simeq 0$.

To summarize this brief digression on the nucleon spin problem: $a_0$, being
the expectation value of a non conserved current, is allowed to evolve in
$Q^2$, both perturbatively and non perturbatively; there is both a quark and a
gluonic contribution to $a_0$ which, perturbatively, is given in Eq.
(\ref{ga}).
The large $Q^2$ perturbative evolution of $a_0$  is computed to be small
($\cal O(\alpha_s^2)$), but the $\alpha_s \Delta g$ contribution does not
vanish. The small $Q^2$ non perturbative evolution may be large and its
description requires non perturbative models. The relative amount
of non perturbative and $\Delta g$ contributions is still an open question;
in this respect any measurements of the gluon spin would be of great
importance.

\vskip 6pt
\nd
{\bf Conclusions}
\vskip 4pt
Needless to say spin effects are fascinating, but tough to understand;
in high energy hadronic interactions both the perturbative, short distance
physics and the non perturbative, large distance one, are interrelated in
subtle and sometime mysterious ways. In order to make successful predictions
or even to explain the existing data, we need a knowledge of hadron wave
functions, hadronization mechanisms, quark-gluon correlations, and so on,
in addition to the usual QCD dynamics. More, detailed and precise spin data
are crucial in allowing the work on spin effects to continue; it might be
a long and hard work but the eventual outcome -- a real understanding of the
hadron structure -- is certainly worth the effort.

\vskip 6pt
\nd
{\bf Acknowledgements}
\vskip 4pt
I would like to thank the organizers of the Conference and Gerry Bunce for
his work and help in the organization of the Session on
``Strong interactions at high energy''.


\vskip 6pt
\baselineskip=6pt
\small
\begin{thebibliography}{99}
\setlength{\parskip}{6pt}
\bibitem{ssa}
\vskip-8pt
Cameron,~P.R., {\it et al.}, \PR{D32}, 3070 (1985).
\bibitem{dsa}
\vskip-8pt
Crabb,~D., {\it et al.}, \PRL{41}, 1257 (1978); Crosbie,~E.A., {\it et al.},
\PR{D23}, 600 (1981); Auer,~I.P., {\it et al.}, \PRL{52}, 808 (1984).
\bibitem{bl}
\vskip-8pt
See, {\it e.g.}, Brodsky,~S.J., and Lepage,~G.P., in {\it Perturbative
Quantum Chromodynamics}, A.H.~Mueller Editor, World Scientific (1989).
\bibitem{for}
\vskip-8pt
Anselmino,~M and Forte,~S., \PRL{71}, 223 (1993).
\bibitem{pias}
\vskip-8pt
Adams,~D.L., {\it et al.}, \ZP{C56}, 181 (1992); \PL{B264} (1991) 462;
\PL{B276} (1992) 531.
\bibitem{siv}
\vskip-8pt
Sivers, D., \PR{D41}, 83 (1990); \PR{D43}, 261 (1991).
\bibitem{qiu}
\vskip-8pt
Qiu, J., and Sterman, G., \PRL{67}, 2264 (1991).
\bibitem{meng}
\vskip-8pt
Boros,~C., Liang,~Z., and Meng,~T., \PRL{70}, 1751 (1993); see also the
talk by T.~Meng in these Proceedings.
\bibitem{szwed}
\vskip-8pt
Szwed,~J., \PL{B105}, 403 (1981).
\bibitem{efre}
\vskip-8pt
Efremov,~A.V. and Teryaev,~O.V., {\it Yad. Fiz.} {\bf 36},
242 (1982) [\SNP{36}, 140 (1982)].
\bibitem{col}
\vskip-8pt
Collins,~J., \NP{B396}, 161 (1993).
\bibitem{artru}
\vskip-8pt
Artru,~X., Czyzewski,~J., and Yabuki,~H.,
{\it Single spin asymmetry in inclusive pion production,
Collins effect and the string model},
preprint LYCEN/9423 and TPJU 12/94, May 1994.
\bibitem{noi}
\vskip-8pt
Anselmino, M., Boglione, M.E., and Murgia, F.,
{\it Single spin asymmetry for $p^{\uparrow}+p\to \pi+X$ in perturbative QCD},
preprint DFTT48/94 and INFNCA-TH-94-27, in preparation.
\bibitem{lam}
\vskip-8pt
Andersson.~B., Gustafson,~G., and Ingelman,~G., \PL{85B}, 417 (1979);
De Grand,~T.A., and Miettinen,~H., \PR{D24}, 2419 (1981).
\bibitem{sof}
\vskip-8pt
Soffer,~J., and T\"ornqvist,~N.A., \PRL{68}, 907 (1992).
\bibitem{pre}
\vskip-8pt
Barni,~R., Preparata,~G., and Ratcliffe,~P.G., \PL{B296}, 251 (1992).
\bibitem{rev}
\vskip-8pt
Anselmino,~M., Efremov,~A., and Leader,~E., preprint CERN-TH/7216/94, to
appear in {\it Phys. Rep}.
\bibitem{emc}
\vskip-8pt
Ashman,~J., {\it et al}, \NP{B328}, 1 (1989).
\end{thebibliography}
\end{document}